\documentstyle[12pt]{article}
\begin{document}
\title{\vspace*{-2.5cm}
Joint description of weak radiative and nonleptonic
hyperon decays in broken SU(3)}
\author{
{P. \.Zenczykowski}$^*$\\
{\em Dept. of Theoretical Physics},\\
{\em Institute of Nuclear Physics, Polish Academy of Sciences}\\
{\em Radzikowskiego 152,
31-342 Krak\'ow, Poland}\\
}
\maketitle
\begin{abstract}
We give a joint description of weak radiative (WR) and 
nonleptonic (NL) hyperon decays (HD) in broken SU(3). 
The two groups of decays are linked via $SU(2)_W$ spin symmetry 
and vector meson dominance (VMD). 
We use experimental information on the parity-conserving (p.c.)
NLHD amplitudes to fix the corresponding WRHD amplitudes.
With the latter known, 
the data on the WRHD branching ratios and
asymmetries permit us to determine the parity-violating (p.v.) WRHD amplitudes
in terms of two parameters corresponding to the two-quark and single-quark
transitions.
We obtain a good description of the data,
and in particular a large $\Sigma ^+ \to p \gamma$ asymmetry. 
Then, using the $SU(2)_W$+VMD route we 
determine the non-soft-meson correction term in the p.v. NLHD amplitudes. 
The latter is shown to subtract a substantial amount 
from the current-algebra commutator thus
leading towards the resolution of the S:P discrepancy in NLHD.

\end{abstract}
\noindent PACS numbers: 11.40.-q, 11.30.Hv, 13.30.-a, 14.20.Jn\\
Keywords: weak radiative nonleptonic hyperon decay\\
\\
$^*$ E-mail:
piotr.zenczykowski@ifj.edu.pl\\
\phantom{$^*$} phone: (48-12)662-8273;
fax: (48-12)662-8458
\newpage

\section{Introduction}

For a long time weak hyperon decays have been presenting us 
with a couple of puzzles (see \cite{DGH,LZ}). These have been in particular:
 the problem of the S:P ratio in the nonleptonic hyperon 
decays (NLHD)
and the issue of a large negative asymmetry in the 
$\Sigma ^+ \to p \gamma $ weak radiative hyperon decay (WRHD).

Both problems emerged several decades ago. The first (S:P) problem is the
inconsistency
between the size of matrix elements of the parity-conserving (p.c.)
Hamiltonian in between ground-state baryon states, as estimated from the 
p.c. (P-wave) NLHD amplitudes, and the size of the same matrix elements 
when estimated from the parity-violating (p.v., S-wave)
NLHD amplitudes via PCAC in the standard soft-pion limit.
The two estimates differ by a factor of around 2 or a little bit larger,
depending on the details of the models used.

The second problem emerged when first experiments hinted that the
$\Sigma^+ \to p \gamma $ asymmetry is large \cite{Gershwin1969}.
Large size of this asymmetry was unexpected since a theorem proved by Hara
\cite{Hara}
stated that in the SU(3) limit the relevant parity-violating amplitude
should vanish. For broken SU(3), having in mind the size of hadron-level
SU(3)-breaking effects elsewhere, one would expect this asymmetry to be
of the order $\pm 0.2$, and not of the order of $-1$ (the present experimental
number is $-0.76 \pm 0.08$).
The situation was further confounded by a number of theoretical calculations
which violated Hara's theorem (even) in the SU(3) limit (see ref. \cite{LZ}).

Some time ago it was pointed out \cite{LZ} that 
the status of Hara's theorem can be clarified through the measurement of
the $\Xi ^0 \to \Lambda \gamma $ decay asymmetry.
By yielding 
a large and negative value of $ -0.78 \pm 0.19 $
for this asymmetry, the
recent NA48 experiment \cite{Schmidt,NA482004} has
decided very clearly in favour of the theorem.

The experimental result of the NA48 collaboration permits to
conclude that theoretical results which violate Hara's theorem in the SU(3)
limit  
constitute artefacts of the relevant approaches.
This concerns both the quark-level calculations of Kamal and Riazuddin
\cite{KR},
and the hadron-level calculation of the present author \cite{Zen89}.
However, the origins of the artefacts are different in these two approaches.

In the quark model of ref. \cite{KR} Hara's theorem is violated
because in the calculations the
intermediate photon-emitting quark enters its mass shell. 
Thus, this quark is treated as a free ordinary particle. 
This leads to a nonvanishing
{\em non-local} contribution and
violates Hara's implicit
assumption that the relevant transition be described in a language of 
a {\em local} hadron-level theory \cite{Zen01}.

The result of ref. \cite{Zen89} follows from ref. \cite{DDH} when
the description of
weak p.v. couplings of vector mesons $V$ to baryons $B$
provided by \cite{DDH} is supplied with
the idea of vector meson dominance (VMD).
In ref. \cite{DDH} the $B'\to VB$ weak p.v.  amplitudes 
are obtained by the application
of $SU(6)_W$ to the {\em full} $B'\to PB$  weak p.v.  amplitudes 
($P$ - pseudoscalar mesons), 
with the latter determined from experimental data on nonleptonic hyperon
decays.
If the applicability of VMD is accepted, the {\em way} in which the $SU(6)_W$-related
$B' \to VB$ counterparts of the $B' \to PB$ amplitudes 
are determined in \cite{DDH} 
must be incorrect.

In fact, ref. \cite{DDH} considers contributions to the p.v. NLHD amplitudes
coming from the current-algebra (CA) commutator term only. 
It is the application of $SU(2)_W$ spin symmetry
to this contribution which ultimately leads to terms violating Hara's theorem.
In general, however, the p.v. NLHD amplitudes contain two terms:
the CA commutator term and the correction term  
which should vanish in the soft-pion limit.
If the latter term is not small for physical pion momentum, 
then its $SU(2)_W$-related
counterpart in WRHD is not small either and could be important
in the description of WRHD.
The observed sign and size of the $\Xi ^0 \to \Lambda \gamma $ asymmetry 
permits us
to make definite conclusions concerning not only Hara's theorem,
but also - via the $SU(2)_W$+VMD route - 
the size and sign of the non-soft-pion term 
in nonleptonic hyperon decays.
As observed in the approach in \cite{Zen03}, in which 
in the parity-violating sector the SU(3) symmetry was exact,
WRHD permit us to establish that the correction term subtracts a substantial part
from the CA commutator contribution, thus working towards the resolution
of the old $S:P$ problem in NLHD.

In the present paper we introduce explicit SU(3) breaking into 
the parity-violating sector of 
the scheme of $\cite{Zen03}$, and show that
despite the fact that the p.v. $\Sigma ^+ \to p \gamma$ amplitude vanishes 
in exact SU(3), in broken SU(3)
this amplitude is comparable in size  to other SU(3)-unsuppressed
p.v. WRHD amplitudes. 
As a result we obtain a large $\Sigma^+ \to p \gamma $  asymmetry.
Our description of the branching ratios and asymmetries in weak radiative
hyperon
decays is in good agreement with the data. Although it deviates
from the experimental data more than the corresponding description of NLHD,
it reproduces both the large size of all observed asymmetries, and provides a
fair description of the branching ratios. In addition,
it predicts a substantial positive asymmetry 
in the $\Xi^- \to \Sigma ^- \gamma$ decay. 
We also show that when SU(3) is broken in the parity-violating sector, 
then the
non-soft-pion contribution to NLHD (obtained from WRHD via the
$SU(2)_W$+VMD route) is of proper sign and 
order of magnitude 
to resolve the $S:P$ problem.

\section{General}

If we write the effective Lagrangian
for nonleptonic hyperon decay $B_i \to B_f \pi $
as
\begin{equation}
\bar{u}_f(A+B\gamma_5) u_i~\Phi_{\pi},
\end{equation} 
where $A$ ($B$) denotes the parity-violating (parity-conserving) amplitude,
the decay rate
 is given by
 \begin{equation}
 \Gamma = \frac{1}{4\pi}\frac{k_{\pi}}{m_i}(E_f+m_f)
 \left[|A|^2+|\bar{B}|^2\right],
 \end{equation}
 where $E_f,m_f$($m_i$) are energy and mass for the final (initial) baryon, 
  $k_{\pi}$ is pion momentum, and
 \begin{equation}
 \bar{B}=\sqrt{\frac{E_f-m_f}{E_f+m_f}}B.
 \end{equation}
The asymmetry is
\begin{equation}
\alpha = \frac{2{\rm{Re}} ~(A^*\bar{B})}{|A|^2+|\bar{B}|^2}.
\end{equation}

Similarly, if the effective Lagrangian for weak radiative hyperon decay
$B_i \to B_f \gamma $ is written as
\begin{equation}
\bar{u}_fi\sigma_{\mu\nu}(p_f-p_i)^{\nu}(C+D\gamma_5)u_i~A^{\mu},
\end{equation}
with $C$ ($D$) being the parity-conserving (violating) amplitude,
then the decay rate is given by
\begin{equation}
\Gamma = \frac{1}{\pi} \left( 
\frac{m^2_i-m^2_f}{2m_i}
\right)^3\left[|C|^2+|D|^2\right],
\end{equation}
and the asymmetry is
\begin{equation}
\alpha = \frac{2{\rm{Re}} ~(C^*D)}{|C|^2+|D|^2}.
\end{equation}
Theoretical models of hyperon decays may relate some or all of the
four amplitudes $A,B,C,D$.
We start with the parity-conserving sector and the relation between 
amplitudes $B$ and $C$.

\section{Parity-conserving amplitudes}

The parity-conserving NLHD amplitudes 
are known to
be well described by the
pole model with the ground-state 
$(56, 1/2^+)$ baryons in the intermediate state.
By $SU(2)_W$ spin symmetry one expects that the same model 
(supplied with the VMD assumption) is adequate for the
description of the p.c. WRHD amplitudes. In this section we present our
version of this approach.

\subsection{Nonleptonic decays}

In the ground-state baryon pole model
the explicit dependence of the p.c. NLHD  amplitudes $B(B_i \to B_f \pi )$ on 
1) $F/D$ describing the SU(3) structure of $\pi B B'$ couplings, and 2) $f_P/d_P$
characterizing the $SU(3)$ structure of the matrix elements of the 
parity-conserving part $H^{p.c.}_W$ of the weak Hamiltonian is (see e.g. \cite{Zen89,Zen94}):
\begin{eqnarray}
B(\Sigma^+ \to p \pi^0) \equiv B(\Sigma^+_0)&=&\frac{1}{\sqrt{2}}\left(\frac{f_P}{d_P}-1\right)
\left(1-\frac{F}{D}\right)N,\nonumber\\
B(\Sigma ^+ \to n \pi ^+) \equiv B(\Sigma^+_+)&=&-\frac{4}{3}N\nonumber\\
B(\Sigma ^- \to n \pi ^-) \equiv B(\Sigma^-_-)&=&\left[\left(\frac{f_P}{d_P}-1\right)\frac{F}{D}-
\frac{1}{3}\left(3\frac{f_P}{d_P}+1\right)
\right]N\nonumber\\
\label{pcNLHD}
B(\Lambda \to n \pi ^-) \equiv B(\Lambda^0_-)&=&-\sqrt{2}B(\Lambda^0_0)\\
&=&\frac{1}{\sqrt{6}}\left[\frac{f_P}{d_P}+3+
\left(3\frac{f_P}{d_P}+1\right)\frac{F}{D}
\right]N\nonumber\\
B(\Xi^- \to \Lambda \pi ^-) \equiv B(\Xi^-_-)&=&-\sqrt{2}B(\Xi^0_0)\nonumber\\
&=&-\frac{1}{\sqrt{6}}\left[3-\frac{f_P}{d_P}+
\left(3\frac{f_P}{d_P}-1\right)\frac{F}{D}
\right]N,\nonumber
\end{eqnarray}
with the standard notation given in the second column (see also 
Table \ref{tablepvNLHDSU6} below).

In writing Eqs(\ref{pcNLHD}) we assumed as
 in
\cite{Zen89,Zen94} that all pole denominators are equal i.e. that
\begin{equation}
\label{equalsplit}
\frac{1}{m_{\Sigma}-m_N}=\frac{1}{m_{\Lambda}-m_N}=\frac{1}{m_{\Xi}-m_{\Sigma}}=
\frac{1}{m_{\Xi}-m_{\Lambda}}\equiv \frac{1}{\Delta m_s},
\end{equation}
with $\Delta m_s = 190~MeV$,
and absorbed them into an overall normalization factor $N$
\begin{equation}
\label{pcNormfactor}
N=\frac{2 m_8}{F_{\pi}}\frac{D~d_P}{\Delta m_s},
\end{equation}
where 
$F_{\pi}=94~MeV$, $m_8$ is some average value of baryon ground-state octet
masses, taken as $m_8=(m_N+m_{\Xi})/2\approx 1130~MeV$,
% $g=13.55$ is the strong coupling constant, $g_A=F+D=1.25$ is the axial
% vector coupling, 
and $d_P$ together with $f_P$ describe the SU(3) structure of
the parity-conserving weak Hamiltonian. 
The form of SU(3) breaking specified in Eq. (\ref{equalsplit})
was used in previous papers on the subject, and 
specifically in ref. \cite{Zen03}, and does not constitute the novelty.
The difference with respect to ref. \cite{Zen03} lies in the explicit
consideration of SU(3) breaking in the parity-violating sector (Section 4).

Our assumption of equal pole denominators (i.e. no $\Sigma - \Lambda$ splitting)
corresponds to the simplest form of $SU(3)_F$ symmetry breaking one can
consider, the whole effect of SU(3) breaking being due to a heavier mass of the
strange quark.  Other elements of the description (such as the strong $B'BP$ 
couplings, or the matrix elements of the p.c. weak Hamiltonian) are 
$SU(3)_F$ - symmetric. 

As the
$\Sigma - \Lambda $ splitting results from spin-spin effects it follows that
taking this splitting into account
would require the consideration of the
influence of spin-spin SU(3) breaking effects
in strong meson-baryon couplings (and possibly in weak transition amplitudes). 
These are not understood well, however. Consequently, $SU(3)_F$ symmetry
was assumed in this paper for the meson-baryon couplings.
For reasons of consistency, therefore, 
we cannot take the spin-interaction-induced $\Sigma-\Lambda$ splitting
into consideration.
Eqs (\ref{pcNLHD}) may be also viewed as just a simple parametrization of 
the $B_i \to B_f \pi$ amplitudes. Transition from these amplitudes
to the amplitudes with pion replaced by
a $U$-spin singlet vector meson $U^0$
(a linear combination of $\rho$, $\omega$, and $\phi$),
 as needed in the next subsection, is achieved
via $SU(6)_W$ symmetry.
If the $B(B_i \to B_f \pi)$ amplitudes are well described 
by Eqs (\ref{pcNLHD}) (and they indeed are,
see Table \ref{tablepc}), then the $B(B_i \to B_fU^0)$ amplitudes should also be
well described.

Our normalization of $f_P$, $d_P$ 
can be read off
from
\begin{equation}
\langle p | H^{p.c.}_w| \Sigma ^+ \rangle = \sqrt{2}~(d_P-f_P),
\end{equation} 
(compare also Table \ref{pvNLHD}).

For $F/D=0.55$ ($F=0.44$, $D=0.81$), $N=-31$ (in units of $10^{-7}$), and 
$f_P/d_P=-1.90$ one obtains a very good description of the data (see Table 1,
also \cite{Zen94}).
Note that our scheme satisfies the $\Delta I=1/2$ rule. Consequently,
one cannot expect here a better agreement 
in view of the violation of this
rule: e.g. the $\Delta I=1/2$ relation
 $\sqrt{2}B(\Sigma^+_0)=B(\Sigma^+_-)-B(\Sigma^-_-)$ 
experimentally reads: $37.6\pm 1.8 = 43.8\pm 0.4 $, indicating that the $\Delta
I =3/2$ effects are of the order of 5-10 \%.

From Eq. (\ref{pcNormfactor}) one finds
\begin{eqnarray}
\label{sizeofdP}
d_P&=&\frac{F_{\pi}}{D}~\frac{\Delta m_s}{2~m_8}N
\approx -3.0 \times 10^{-5}~MeV\nonumber\\
f_P&\approx & 5.8 \times 10^{-5}~MeV.
\end{eqnarray}

Our $f_P$ and $d_P$ parameters are related to the ones used in \cite{DGH}
by
\begin{eqnarray}
f_P&=&-2\sqrt{3} F_{\pi}f_P(\cite{DGH})\\
d_P&=&-2\sqrt{3} F_{\pi}d_P(\cite{DGH}).
\end{eqnarray}
The values of $f_P(\cite{DGH})=-1.44 \times 10^{-7}$ and 
$d_P(\cite{DGH})=0.8 \times 10^{-7}$ given in \cite{DGH} 
correspond to our
\begin{eqnarray}
d_P&\approx & -2.6 \times 10^{-5}~MeV \nonumber\\
\label{DGHfPdP}
f_P&\approx & 4.7 \times 10^{-5}~MeV.
\end{eqnarray} 
The difference between the latter numbers and the estimates of
Eq.(\ref{sizeofdP}) indicates how large the uncertainty in 
the extracted values of $f_P$ and $d_P$ might be \footnotemark[1].
\footnotetext[1]{ Uncertainties of this order might result e.g. 
from the treatment of kaon
poles which were neglected by us but  taken into account 
in \cite{DGH}.}

\begin{table}
\caption{P-wave NLHD amplitudes $B(B_i \to B_f \pi)$ 
(in units of $10^{-7}$) using
Eq.(\ref{pcNLHD}) (with $f_P/d_P=-1.9$, $F/D=0.55$, $N=-31 \times 10^{-7}$) 
and the data }
\label{tablepc}
\begin{center}
\begin{tabular}{ccc}
\hline
Decay          & Eq.(\ref{pcNLHD})               & Data     \\
\hline
$\Sigma ^+_0 $ & $28.6$ & $26.6  \pm 1.3$   \\
$\Sigma ^+_+ $ & $41.3$  & $42.4  \pm 0.35$  \\
$\Sigma ^-_- $ & $0.9$                    & $-1.44 \pm 0.17$  \\
$\Lambda^0_- $ & $18.8 $                    & $22.1  \pm 0.5$\\
$\Xi    ^-_- $ & $15.4$          & $16.6  \pm 0.8$\\
\hline
\end{tabular}
\end{center}
\end{table}

\subsection{Radiative decays}

For the WRHD the parity-conserving amplitudes $C(B_i \to B_f \gamma )$,   
obtained in
the ground-state baryon pole model 
from the p.c. NLHD amplitudes via the $SU(2)_W$+VMD route, are
 given by
\begin{equation}
\label{CfromB}
C(B_i \to B_f \gamma) = 
\left(\frac{e}{g}\right)\frac{1}{(m_i+m_f)\sqrt{2}}~B(B_i \to B_f U^0 ).
\end{equation}
In the above equation 
$e/g =0.0606$ is the VMD factor ($e^2/4\pi=1/137$, $g=5.0$), and
$B(B_i \to B_f U^0)$ describe amplitudes for the emission of a linear
superposition $U^0$ of virtual vector mesons $\rho ^0, \omega, \phi$,
corresponding to a photon and obtained by the $SU(6)_W$ symmetry
from the NLHD amplitudes $B_i \to B_f \pi$ of Eq. (\ref{pcNLHD}).
For the $B(B_i \to B_f U^0)$ amplitudes one gets:
\begin{eqnarray}
B(\Sigma^+ \to p~U^0)& = & \sqrt{2}\left(\frac{f_P}{d_P}-1\right)
(\mu_{\Sigma^+}-\mu_{p})\frac{N}{\mu _p D}\nonumber\\
B(\Sigma^0 \to n~U^0)& = &\left[-\left(\frac{f_P}{d_P}-1\right)
(\mu _{\Sigma ^0}-\mu _{n})+\frac{1}{\sqrt{3}}
\left(3 \frac{f_P}{d_P}+1\right)\mu _{\Sigma \Lambda}
\right]\frac{N}{\mu _p D}\nonumber \\
\label{pcWRHD}
B(\Lambda \to n U^0)&=&\left[\frac{1}{\sqrt{3}}
\left(3 \frac{f_P}{d_P}+1\right)(\mu _{\Lambda}-\mu _{n})
-\left(\frac{f_P}{d_P}-1\right)
\mu _{\Sigma \Lambda}
\right]\frac{N}{\mu _p D}\\
B(\Xi^0\to\Lambda U^0)&=&\left[-\frac{1}{\sqrt{3}}
\left(3 \frac{f_P}{d_P}-1\right)(\mu _{\Xi^0}-\mu _{\Lambda})
-\left(\frac{f_P}{d_P}+1\right)
\mu _{\Sigma \Lambda}
\right]\frac{N}{\mu _p D}\nonumber\\
B(\Xi^0 \to \Sigma^0 U^0)&=&\left[\left(\frac{f_P}{d_P}+1\right)
(\mu _{\Xi ^0}-\mu _{\Sigma ^0})+\frac{1}{\sqrt{3}}
\left(3 \frac{f_P}{d_P}-1\right)\mu _{\Sigma \Lambda}
\right]\frac{N}{\mu _p D}\nonumber\\
B(\Xi^-\to\Sigma^- U^0)&=&-\sqrt{2}\left(\frac{f_P}{d_P}+1\right)
(\mu _{\Xi^-}-\mu _{\Sigma ^-})\frac{N}{\mu _p D},\nonumber
\end{eqnarray}
where we used equal mass splittings as suggested both by the success of
Eqs (\ref{pcNLHD}) when describing the data, 
and by the analysis of the Lee-Sugawara relations
performed in \cite{Zen91}. The appearance of magnetic moments 
in Eqs (\ref{pcWRHD}) will be explained
shortly.

The particular form of the r.h.s. of Eqs (\ref{pcWRHD}) was obtained in
\cite{Zen91}, where the standard expressions for the $B(B_i \to B_f U^0)$ 
amplitudes (depending on the $F$ and $D$ couplings
and similar to Eqs (\ref{pcNLHD})) were rewritten in terms
of the corresponding magnetic moments to which they would be proportional
in the $SU(6)$ symmetry limit.
The reason for using this representation is that ultimately we want to
describe photon couplings which 
originate from the 
$\bar{u}_1\sigma _{\mu\nu}q^{\mu}u_2A^{\nu}$ terms
and thus are expressed 
in terms of the anomalous parts of baryon
magnetic moments.
Now,
the description of baryon magnetic moments provided by $SU(6)$
(or when the strange quark is assumed to be heavier and has a smaller magnetic
moment) is not good enough for our purposes. This is because there are
substantial cancellations between various terms in Eqs (\ref{pcWRHD}), and the
results depend on the detailed values of baryon magnetic moments.
In fact, it is known that substantial nonadditivities are observed in the
experimental values of baryon magnetic moments: a thorough analysis performed in
\cite{Lipkin} revealed that the nonstrange quark contributions 
to baryon magnetic
moments in proton and neutron are significantly larger than in the baryons
containing strange quarks. 
Since ultimately we want to describe photon couplings, the need to use magnetic
moments on the l.h.s. of Eq. (\ref{CfromB}) requires that its r.h.s. takes
them into account as well, as in Eqs (\ref{pcWRHD}), thus modifying the
vector meson couplings accordingly.

In order to describe the photon couplings best,
we chose to use in Eqs (\ref{pcWRHD}) 
the experimental values themselves, i.e. (from \cite{PDG}):
$\mu_p=2.793$, $\mu _n=-1.913$, $\mu_{\Sigma^+}=2.46\pm 0.01$, 
$\mu _{\Sigma \Lambda}= 1.61\pm 0.08$, $\mu _{\Sigma^-}=-1.16\pm 0.025 $,
$\mu _{\Lambda}=-0.613\pm0.004$, $\mu _{\Xi^0}=-1.25 \pm 0.014$, 
$\mu _{\Xi ^-}=-0.651\pm 0.003$, 
$\mu _{\Sigma ^0}=(\mu _{\Sigma ^+}+\mu _{\Sigma^-})/2$.
From Eq.(\ref{CfromB}), using the fit of Table \ref{tablepc}, 
one can predict the
p.c. WRHD amplitudes $C(B_i \to B_f \gamma)$. 
The relevant numbers for the related $B(B_i \to B_f U^0)$ amplitudes
are given in Table \ref{tablepcWRHD}.

The numbers given in Table \ref{tablepcWRHD} result from
cancellations between various terms in Eqs (\ref{pcWRHD}). 
Such cancellations are strongest for the $B(\Lambda \to n U^0 )$ amplitude.
Specifically, for $B(\Lambda \to n U^0 )$
the three terms seen in Eqs (\ref{pcWRHD}) contribute as follows:
the $\mu _{\Lambda}$ term gives $-22.9$, the $\mu _n$ term: $+71.4$, and
the $\mu _{\Sigma \Lambda}$ term: $-64.2$. A change of $\mu
_{\Sigma \Lambda}$ by one standard deviation from $1.61$ to
$1.69$ leads to the absolute value 
of $B(\Lambda \to n U^0 )$ larger 
 than the value given in Table \ref{tablepcWRHD} by
 20\%. 
 For the remaining amplitudes
the uncertainty in 
$\mu _{\Sigma \Lambda}$ leads to errors of the order of a few percent only.

One should keep also
in mind that 
our approach leads to an additional uncertainty in the size of the p.v.
amplitudes
for the $\Sigma^+ \to p \gamma $ and 
$\Xi ^- \to \Sigma ^- \gamma $ decays.
This is because 
within our treatment of SU(3) breaking the differences of the 
anomalous parts of baryon magnetic moments reduce to 
the differences of baryon magnetic moments themselves.

\begin{table}
\caption{P-wave amplitudes $B(B_i \to B_f U^0)$ 
(in units of $ 10^{-7}$) as obtained from Eqs (\ref{pcWRHD}) with $f_P/d_P=-1.9$,
$d_P=-3.0\times 10^{-5}~MeV$, $N=-31 \times 10^{-7}$.}
\label{tablepcWRHD}
\begin{center}
\begin{tabular}{cc}
\hline
Decay          & Eq.(\ref{pcWRHD})     \\
\hline
$\Sigma ^+ \to p U^0    $ & $-18.8$   \\
$\Sigma ^0 \to n U^0    $ & $-42.1$   \\
$\Lambda \to n U^0      $ & $-15.7$   \\
$\Xi^0 \to \Lambda U^0  $ & $+13.9$   \\
$\Xi^0 \to \Sigma^0 U^0 $ & $+62.1$   \\
$\Xi^- \to \Sigma ^- U^0$ & $-8.9$   \\
\hline
\end{tabular}
\end{center}
\end{table}

\section{Parity-violating amplitudes}

\subsection{Nonleptonic decays}

In the first approximation the parity-violating 
amplitudes $A(\alpha \to \beta
\pi)$ are given
by the soft-meson estimate \cite{GM,AdlerDashen}:
\begin{eqnarray}
\label{softmeson}
\langle \pi _a \beta | H_W^{p.v.} | \alpha \rangle & = &\frac{-i}{F_{\pi}}
\langle \beta |[F^5_a,H_W^{p.v.}]|\alpha \rangle + q _{\mu} M^{\mu}_a \nonumber \\
&\stackrel{q \to 0}{\longrightarrow} &  \frac{-i}{F_{\pi}}
\langle \beta |[F^5_a,H_W^{p.v.}]|\alpha \rangle,
\end{eqnarray}
where $F_5^a$ is the axial charge,
$F_{\pi}=94~MeV$, and $H^{p.v.}_W$ is the p.v. part of the weak Hamiltonian.
Since
\begin{equation}
[F^5_a,H_W^{p.v.}]=-[F_a,H_W^{p.c.}],
\end{equation}
the commutator term $\langle \beta| [F^5_a,H_W^{p.v.}]| \alpha \rangle $ 
may be expressed in terms of 
$\langle \beta| [F_a,H_W^{p.c.}]| \alpha \rangle $, with $F_a$ the generator of
ordinary flavour symmetry.
Consequently, the commutator term may be
expressed in terms of the
matrix elements of $H^{p.c.}_W$ between appropriate baryon states.

In Table \ref{pvNLHD} we gather expressions for the p.v. NLHD amplitudes  
given in terms of the
commutator parameters
$f_P$ and $d_P$ as well as in terms of
 the observed p.v. amplitudes $b_S$ and $c_S$ 
corresponding to the contributions of the topology of the
$W$-exchange and penguin diagrams respectively. 
The relevant diagrams are shown in Fig.1.
As shown,
the comparison with the data requires 
\begin{eqnarray}
b_S &\approx & -5 \times 10^{-7}  \nonumber \\
\label{bScSdata}
c_S &\approx & +12 \times 10^{-7}.
\end{eqnarray}
(Diagrams (a) and (a') of Fig.1 do not contribute in NLHD.)
 
 When the SU(3) amplitudes $f$ and $d$ describing the p.c. transitions are
 extracted from 
 $b_S$
 and $c_S$ one obtains:
\begin{eqnarray}
f_S&\equiv &-\frac{F_{\pi}}{4}(b_S-\frac{2}{3}c_S)
\approx 3.1 \times 10^{-5} ~MeV\nonumber \\
d_S&\equiv &\frac{F_{\pi}}{4}b_S \approx - 1.2 \times
10^{-5}~MeV.\footnotemark[2]
\end{eqnarray}   
\footnotetext[2]{The above values are in full 
agreement with those given in Eq.(6.12) of \cite{DGH},
with the relative relation being
$f_S =-2\sqrt{3}  F_{\pi} f_S(\cite{DGH})$,
 $d_S =-2\sqrt{3}  F_{\pi} d_S(\cite{DGH})$, where
  $f_S(\cite{DGH})=-0.92 \times 10^{-7}$,
 $d_S(\cite{DGH})=0.38 \times 10^{-7}$.}

Comparing with Eq.(\ref{sizeofdP}) we see that
\begin{eqnarray}
d_P & \approx & 2.6 ~d_S\nonumber \\
\label{factor2}
f_P & \approx & 1.9 ~f_S.
\end{eqnarray}
If one introduces
\begin{eqnarray}
b_P&\equiv &\frac{4}{F_{\pi}}d_P\nonumber \\
c_P&\equiv &\frac{6}{F_{\pi}}(f_P+d_P),
\end{eqnarray}
one further finds 
\begin{eqnarray}
b_P&\approx & -12.9 \times 10^{-7}\nonumber \\
\label{bPcPdata}
c_P&\approx & 17.5 \times 10^{-7}.
\end{eqnarray}
If instead of Eq.(\ref{sizeofdP}) one uses the estimates of $f_P$ and $d_P$ given
in Eq.(\ref{DGHfPdP}) one obtains
\begin{eqnarray}
d_P & \approx & 2.2 ~d_S\nonumber \\
\label{factortwo}
f_P & \approx & 1.5 ~f_S,
\end{eqnarray}
and
\begin{eqnarray}
b_P&\approx & -11.1 \times 10^{-7}\nonumber \\
\label{bPcPdata2}
c_P&\approx & 13.4 \times 10^{-7}.
\end{eqnarray}

Thus, the description of NLHD amplitudes in terms of the leading order terms:
the ground-state-baryon pole contribution for the P-waves,
and the current-algebra commutator term for the S-waves presents a problem.
While soft-meson theorems lead to $f_P=f_S$ and $d_P=d_S$,
the values of $f_P$ and $d_P$ as extracted from the parity-conserving
amplitudes are larger by a factor of around 2 than those (i.e. $f_S,d_S$)
needed for the description of the parity-violating amplitudes (the S:P problem). 
In addition, their ratios
differ significantly:
$f_P/d_P\approx -1.8 ~{\rm to}~ -1.9$, while $f_S/d_S\approx -2.6$.
When the problem is expressed
in terms of amplitudes $b$ and $c$ corresponding to $W$-exchange-mediated
and single-quark transitions we observe from Eqs
(\ref{bScSdata},\ref{bPcPdata},\ref{bPcPdata2})
that the reduction of $b_P$ to $b_S$ is much stronger than 
a similar reduction of $c_P$ to $c_S$.
This indicates that the dominant correction to the soft-meson formula
originates from the $W$-exchange diagrams.

Following the success of the ground-state baryon pole model in the p.c. sector
it was proposed in \cite{LeY} that the soft-meson expression for the p.v.
amplitudes should be supplemented with a substantial  
correction term $R = q_{\mu}M^{\mu}$
originating
from a pole model contribution of the negative-parity 
intermediate excited states ($(70,1/2^-)$ in the quark model). 

\begin{table}
\caption{Parity violating amplitudes in NLHD}
\label{pvNLHD}
\begin{center}
\begin{tabular}{c|cc|rr}
Amplitude&\multicolumn{2}{c|}{Parametrization}&\multicolumn{2}{c}{Values (in
units of $10^{-7}$)}\\
 & Commutator & Diagram  & Experiment & Description \\
 &&decomposition&&($b_S=-5$, \\
 &&&&$c_S=12$)\\
\hline
$A(\Lambda
^0_-)$&$\frac{1}{2F_{\pi}}\sqrt{\frac{2}{3}}(3f_P+d_P)$&$-\frac{1}{2\sqrt{6}}(b_S-c_S)$&$3.25$&$3.47$\\
$A(\Lambda
^0_0)$&$-\frac{1}{2F_{\pi}}\frac{1}{\sqrt{3}}(3f_P+d_P)$&$\frac{1}{4\sqrt{3}}(b_S-c_S)$&$-2.37$&$-2.46$\\
$A(\Sigma ^+_+)$&$0$&$0$&$0.13$&$0$\\
$A(\Sigma
^+_0)$&$-\frac{1}{2F_{\pi}}\sqrt{2}(f_P-d_P)$&$\frac{1}{2\sqrt{2}}(b_S-\frac{c_S}{3})$&$-3.27$
&$-3.18$\\
$A(\Sigma ^-_-)$&$\frac{1}{2F_{\pi}}2(f_P-d_P)$&$-\frac{1}{2}(b_S-\frac{c_S}{3})$&$4.27$&$4.50$\\
$A(\Xi
^0_0)$&$-\frac{1}{2F_{\pi}}\frac{1}{\sqrt{3}}(-3f_P+d_P)$&$-\frac{1}{2\sqrt{3}}(b_S-\frac{c_S}{2})$&$3.43$&$3.18$\\
$A(\Xi
^-_-)$&$\frac{1}{2F_{\pi}}\sqrt{\frac{2}{3}}(-3f_P+d_P)$&
$\frac{1}{\sqrt{6}}(b_S-\frac{c_S}{2})$&$-4.51$&$-4.49$\\
\hline
\end{tabular}
\end{center}
\end{table}

The pole model contribution contains two terms (diagrams (1) and (2) in Fig.1
- for both (b) and (c) type transitions), 
differing in the order
of the action of weak and strong transitions.
The parity-violating weak transition is described by (see e.g. \cite{Zen99})
\begin{equation}
a_{kl}~\bar{u}_ku_l,
\end{equation}
where the pair of indices $k$, $l$ describes a pair of $1/2^+$ and $1/2^-$
baryons $(B,B^*)$, i.e. $(k,l) \leftrightarrow (B^*_k,B_l)$ or $(B_k,B^*_l)$.
Hermiticity and CP invariance require $a$ to be purely imaginary and
antisymmetric \cite{Zen99,Okun}:
\begin{equation}
a_{kl}=-a_{lk}.
\end{equation}

The (parity-conserving) strong transition is described by a gradient
coupling of the pion. For simplicity, we shall consider $\pi ^0$ only 
as its C-parity is well-defined:
\begin{equation}
\label{strong}
f_{kl}~\bar{u}_k\!\!\not\!qu_l~\pi^0,
\end{equation} 
with $k$, $l$ describing as before a pair of $1/2^+$ and $1/2^-$
baryons.
Hermiticity and CP invariance require $f$ to be real and antisymmetric:
\begin{equation}
f_{kl}=-f_{lk}.
\end{equation}

Diagrams (1) (i.e. (b1) and (c1)) and (2) (i.e. (b2) and (c2)) 
lead  to the following total contribution of the $1/2^-$ poles to the $B_i \to
B_f \pi^0$ decay:
\begin{equation}
\label{totalRinNLHD}
\left\{ \frac{f_{fk^*} a_{k^*i}}{m_i-m_{k*}}
+\frac{a_{fk^*} f_{k^*i}}{m_f-m_{k*}}\right\}
\bar{u}_f\!\!\not\!qu_i~\pi^0,
\end{equation}
where the first (second) term originates from diagram (1) (respectively (2))
and the subscripts $k^*$ label intermediate excited $1/2^-$ baryons. 
When $m_f=m_i$ one finds that the term in braces is symmetric under $i
\leftrightarrow f$ interchange.

Using symmetry properties of $a_{k^*l}$ and $f_{k^*l}$ 
one can see from Eq.(\ref{totalRinNLHD}) that for $m_i\approx m_f$
the contribution from diagram (1) (alternatively diagram (2)) in 
$B_i \to B_f \pi^0$ transition
must be equal to the contribution from diagram (2) 
(alternatively diagram (1)) in
$B_f \to B_i \pi^0$ transition.

In the present paper the sums over intermediate $(70,1/2^-)$ baryons $k^*$ are 
not actually performed since our approach deals with their end results only. 
Thus, only the $SU(6)_W$ structure of the latter is important.
Originally, calculations of this structure were
carried out in \cite{Zen89,DDH}.  
The only difference of this paper with respect
to \cite{Zen89,DDH} is a different relative sign 
of contributions 
from diagrams
(1) and (2). 
The presence of this difference is understable 
as $SU(6)_W$ relates all amplitudes
corresponding to
diagram (1) (alternatively diagram (2)),
but does not relate amplitudes of diagrams (1) to those of diagrams (2).
Relation between amplitudes corresponding
 to diagrams (1) and (2) is dictated by the considerations above
 and in particular by the gradient form of pion coupling.
In the $SU(6)_W$ symmetric scheme supplied with the above $i \leftrightarrow f$
symmetry
condition the relevant expressions may be therefore
readily copied from \cite{Zen94,Zen99} with appropriate sign adjustments.
These amplitudes, expressed in terms of amplitudes $b_R$ and $c_R$,
corresponding to $W$-exchange and penguin diagrams respectively, are
gathered in Table \ref{tablepvNLHDSU6}. 
In order to show in an explicit way the 
$i \leftrightarrow f$ 
symmetry property required by the gradient coupling,
in addition to the amplitudes for 
the observed decays ($\Sigma ^+_{0}$, $\Sigma ^+_{+}$ etc.)
 we also listed 
the amplitudes for
the kinematically forbidden transitions $p \to \Sigma ^+ \pi ^0$, $p \to p
\pi^0$, and $\Sigma^+ \to p U^0_P$ (with $U^0_P=(\sqrt{3}\pi^0+\eta_8)/\sqrt{2}$).

\begin{table}
\caption{Contributions of diagrams (b1), (b2) and (c1), (c2) of Fig.1 
to NLHD amplitudes using $SU(6)_W$ with gradient pion coupling 
(using \cite{Zen94,Zen99}, $U^0_P=(\sqrt{3}\pi^0+\eta_8)/\sqrt{2}$)}
\label{tablepvNLHDSU6}
\begin{center}
\begin{tabular}{cccccc}
&Transition & (b1) & (b2) & (c1) & (c2) \\
\hline
$\Sigma ^+_0 $&$\Sigma ^+ \to p \pi^0 $& $0$ & $\frac{1}{2\sqrt{2}}b_R$&
$\frac{1}{6\sqrt{2}}c_R$& $0$\\
$\Sigma ^+_+ $& $\Sigma ^+ \to n \pi ^+$ & $0$&$0$&$0$&$0$\\
$\Sigma ^-_-$&$\Sigma ^- \to n \pi ^-$& $0$ &$-\frac{1}{2}b_R$&
$-\frac{1}{6}c_R$&$0$\\
$\Lambda ^0_-$&$\Lambda \to p \pi^-$& $0$&$-\frac{1}{2\sqrt{6}}b_R$&
$-\frac{1}{2\sqrt{6}}c_R$& $0$\\
$\Lambda ^0_0$&$\Lambda \to n \pi ^0$& $0$&$\frac{1}{4\sqrt{3}}b_R$&
$\frac{1}{4\sqrt{3}}c_R$&$0$\\
$\Xi ^-_-$& $\Xi ^- \to \Lambda \pi ^-$& $0$& $\frac{1}{\sqrt{6}}b_R$&
$\frac{1}{2\sqrt{6}}c_R$& $0$\\
$\Xi ^0_0$& $\Xi ^0 \to \Lambda \pi ^0$ & $0$ & $-\frac{1}{2\sqrt{3}}b_R$&
$-\frac{1}{4\sqrt{3}}c_R$&$0$\\
&$p \to \Sigma ^+ \pi ^0$ & $\frac{1}{2\sqrt{2}}b_R$& $0$ & $0$ &
$\frac{1}{6\sqrt{2}}c_R$\\
&$p \to p \pi^0$&$\frac{1}{2\sqrt{2}}b_R\cot \theta_C$&
$\frac{1}{2\sqrt{2}}b_R \cot \theta_C$&$\frac{1}{6\sqrt{2}}c_R\cot \theta_C$&
$\frac{1}{6\sqrt{2}}c_R\cot \theta_C$\\
&$\Sigma^+ \to p U^0_P$&$\frac{1}{2\sqrt{6}}b_R$&$\frac{1}{2\sqrt{6}}b_R$
&$\frac{1}{6\sqrt{6}}c_R$&$\frac{1}{6\sqrt{6}}c_R$\\
\hline
\end{tabular}
\end{center}
\end{table}

When expressed in the language of $b$ and $c$ amplitudes 
the  correction term $R$ leads to :
\begin{eqnarray}
b_S&=&b_P+b_R\nonumber\\
\label{bRcR}
c_S&=&c_P-c_R,
\end{eqnarray}
with $b_R$, $c_R$ representing the corrections.
The terms $b_R$ and $c_R$ are proportional to $m_i-m_f$ (in Eq. 
(\ref{totalRinNLHD}) this originates from the $\bar{u}_f\,\!\!\!\not\!q\,u_i$
factor) and therefore they vanish in the limit when $q^0 = m_i-m_f \to 0 $.
The above formulas are quite general as they follow from SU(3) and the gradient-coupling
form only. Later we shall consider SU(3) breaking in the propagators
of the intermediate $1/2^-$ states.
The size of the $1/2^-$-induced correction terms may be estimated 
in a quark model
\cite{LeY} and is sizable.
Still, the error of such an estimate may be substantial 
(in \cite{LeY} it is judged to be of the order of 50\%).
Consequently, we shall try a different route and estimate the size and sign
of the correction term 
from weak radiative hyperon decays using
VMD and $SU(2)_W$ ($SU(6)_W$) spin symmetry.

\subsection{Radiative decays}

The parity-violating WRHD amplitudes are obtained from those of NLHD amplitudes
by
replacing the emission of a pion with that of a photon. 
The connection between the two sets of amplitudes 
may be obtained via VMD and the symmetry of 
$SU(6)_W$ ($SU(2)_W$). 

When VMD and $SU(6)_W$ are together
applied to the commutator term of NLHD they lead to
the violation of
Hara's theorem \cite{Hara} in the SU(3) limit \cite{Zen89}.
Although the simple quark model, the bag model and early experiments also hinted 
at the violation of Hara's
theorem in that limit, thus suggesting that some assumption of the theorem is
violated, the
question of Hara's theorem violation is now experimentally 
settled in the negative by the measurement of the $\Xi^0 \to \Lambda \gamma$ 
asymmetry \cite{NA482004}
as discussed in \cite{Zen03}.
Thus, in agreement with the general theoretical expectations
(c.f. the argument presented in \cite{Zen03}),
the soft-meson commutator term present in the p.v. NLHD amplitudes
has no $SU(2)_W$-generated counterpart in the WRHD sector.

Consequently, up to an appropriate VMD factor, 
the parity-violating WRHD amplitudes are the $SU(2)_W$-generated
counterparts of the $q_{\mu}M^{\mu}$ term in NLHD.
As discussed in the previous subsection and in
\cite{LeY}, in NLHD this term originates from the pole-model contribution of the
intermediate $1/2^-$ excited baryons. Following \cite{LeY}, it
was therefore proposed in \cite{Gav} that the parity-violating WRHD transitions
are generated in an analogous manner.

 Thus, in the considerations of the previous Section one has to replace
 the
strong vertices of Eq.(\ref{strong}) by the electromagnetic ones:
\begin{equation}
\mu _{kl}~ \bar{u}_k i \sigma ^{\mu \nu} \gamma_5 q_{\nu} u_l~ A_{\mu},
\end{equation} 
with $k$, $l$ describing as before a pair of $1/2^+$ and $1/2^-$ baryons.
Hermiticity and CP invariance require $\mu$ to be purely imaginary and
symmetric \cite{Zen99,Okun}:
\begin{equation}
\mu _{kl}=\mu_{lk}.
\end{equation}
Diagrams (1) and (2) lead to the following total contribution of the $1/2^-$
poles to the $B_i \to B_f \gamma $ decay:
\begin{equation}
\label{pvWRHDpole}
\left\{ \frac{\mu
_{fk^*}a_{k^*i}}{m_i-m_{k^*}}+\frac{a_{fk^*}\mu_{k^*i}}{m_f-m_{k^*}}
\right\}
\bar{u}_f i \sigma ^{\mu \nu} \gamma _5 q_{\nu} u_i~ A_{\mu}.
\end{equation}
When $m_i\approx m_f$ one finds that the term in braces in Eq.(\ref{pvWRHDpole})
is antisymmetric under $i \leftrightarrow f$ interchange.

Using symmetry properties of $a_{k^*l}$ and $\mu_{k^*l}$ 
one can see from Eq.(\ref{pvWRHDpole}) that for $m_i\approx m_f$
the contribution from diagram (1) (alternatively diagram (2)) in 
$B_i \to B_f \gamma$ transition
must be opposite in sign and equal in absolute magnitude 
to the contribution from diagram (2) 
(alternatively diagram (1)) in
$B_f \to B_i \gamma$ transition.
For $i=\Sigma^+$ and $f=p$ it then
follows from the symmetry of the weak interaction
Hamiltonian under $s \leftrightarrow d$ (and thus $\Sigma ^+ \leftrightarrow p$)
that the term in braces in Eq.(\ref{pvWRHDpole}) has to be both antisymmetric
nad symmetric under $i \leftrightarrow f$ interchange.
This means that
the p.v. $\Sigma ^+ \to p \gamma$ amplitude   must vanish
in the $SU(3)$ limit (Hara's theorem \cite{Hara}).

As in the case of NLHD, in the $SU(6)_W$ approach to WRHD
 the sums over intermediate $(70,1/2^-)$ baryons $k^*$ are not 
 actually performed, since we deal with the end results only.
  Up to an appropriate normalization and the VMD factor 
 the $SU(6)_W$ scheme relates then the correction terms
 to the p.v. NLHD amplitudes
 and the p.v. WRHD amplitudes 
 for diagrams (1). Similar connection exists (separately) for diagrams (2).
 The obtained amplitudes $A(B_i \to B_f U^0)$, 
 copied from \cite{LZ,Zen94} with appropriate sign
 adjustment (as in Table \ref{tablepvNLHDSU6}), are gathered in Table
 \ref{tablepvWRHDSU6}. Contributions from the coupling of the $U^0$ vector 
 meson (later photon) to the
 strange quark are described by parameter $\epsilon$ ($=1$ in SU(3)). 
 All single-quark contributions  
 may be lumped into a single unknown parameter
$s_R$ (which includes $c_R$ and the amplitudes corresponding to diagrams (a),
(a') of
Fig.1, see \cite{Zen89,DDH}). Any 
$SU(3)$-breaking effects may be absorbed into its definiton.
The relevant contributions are also gathered in Table \ref{tablepvWRHDSU6}.

\begin{table}
\caption{Contributions of diagrams (b1), (b2)  of Fig.1 and single quark
transitions
to $A(B_i \to B_f U^0)$ amplitudes using $SU(6)_W$ and Table \ref{tablepvNLHDSU6} 
(from \cite{LZ,Zen94})}
\label{tablepvWRHDSU6}
\begin{center}
\begin{tabular}{cccc}
Process & diagram (b1) & diagram (b2) & single-quark \\
\hline
$\Sigma ^+ \to p U^0 $ & $\frac{1}{3\sqrt{2}}~b_R$ & -
$\frac{2+\epsilon}{9\sqrt{2}}~b_R$ & $\frac{1}{3\sqrt{2}}~s_R$ \\
$\Sigma ^0 \to n U^0$ & $\frac{1}{6}~b_R$ & $\frac{2+\epsilon}{18}~b_R$ & $-
\frac{1}{6}~s_R$\\
$\Lambda \to n U^0$ & $-\frac{1}{6\sqrt{3}}~b_R$ & 
$\frac{2+\epsilon}{6\sqrt{3}}~b_R $ & $-\frac{\sqrt{3}}{2}~s_R$ \\
$\Xi ^0 \to \Lambda U^0 $ & $0$ & $-\frac{2+\epsilon}{9\sqrt{3}}~b_R$
&$\frac{1}{2\sqrt{3}}~s_R$\\
$\Xi ^0 \to \Sigma ^0 U^0 $& $-\frac{1}{3}~b_R$& $0$ & $ -\frac{5}{6}~s_R$\\
$\Xi ^- \to \Sigma ^- U^0 $&$0$&$0$&$\frac{5}{3\sqrt{2}}~s_R$\\
\end{tabular}
\end{center}
\end{table}

In the pole model with broken SU(3) ($\Delta m_s =190~MeV$)
the contribution from diagrams (1) and (2) given in Tables \ref{tablepvNLHDSU6},
\ref{tablepvWRHDSU6}
will be somewhat modified. 
Namely, diagrams (1) are associated with the presence of mass denominators
$\Delta \omega -\Delta m_s$, while for diagrams (2) these mass denominators
contain $\Delta \omega +\Delta m_s$,
where $\Delta \omega \approx 570~MeV$ is the average splitting between the 
$(56,1/2^+)$ and $(70,1/2^-)$ multiplets.
Assuming that all of the SU(3) breaking originates from mass
differences 
(plus possibly from a reduced coupling of $U^0$/photon to the strange quark, i.e.
$\epsilon <1$),
we may take it into account by multiplying the contributions of diagrams (1)
by 
$\Delta \omega /(\Delta \omega -\Delta m_s) \equiv 1/(1-x)$ with $x \approx
1/3$. For diagrams (2) the relevant factor is $1/(1+x)$.
It is mainly through the presence of these SU(3) breaking effetcs that the
present paper differs from ref. \cite{Zen03}.

Using the above considerations one obtains the following expressions for the  
parity-violating WRHD amplitudes:
\begin{equation}
\label{DintermsofA}
D(B_i\to B_f \gamma )= \left( \frac{e}{g} \right) \frac{1}{(m_i-m_f)\sqrt{2}}
A(B_i \to B_f U^0),
\end{equation}
where amplitudes $A$ are related by $SU(2)_W$ to the (vanishing in the
soft-meson limit) correction terms
in NLHD:
\begin{eqnarray}
A(\Sigma ^+ \to p U^0)& = & 
\frac{1}{9\sqrt{2}}~\frac{6x+(1-\epsilon)(1-x)}{1-x^2}~b_R+
\frac{1}{3\sqrt{2}}~s_R\nonumber \\
A(\Sigma ^0 \to n U^0) &=& \frac{1}{18}~
\frac{6-(1-\epsilon)(1-x)}{1-x^2}~b_R-\frac{1}{6}~s_R\nonumber \\
A(\Lambda \to n U^0) &= &
-\frac{1}{6\sqrt{3}}~\frac{4x-2+(1-\epsilon)(1-x)}{1-x^2}~
b_R-\frac{\sqrt{3}}{2}~s_R\nonumber \\
A(\Xi^0 \to \Lambda U^0)&=&
-\frac{2+\epsilon}{9\sqrt{3}}~\frac{1-x}{1-x^2}~b_R+\frac{1}{2\sqrt{3}}~s_R
\nonumber \\
A(\Xi^0 \to \Sigma ^0 U^0)&=& -\frac{1}{3}~\frac{1+x}{1-x^2}~b_R-
\frac{5}{6}~s_R\nonumber \\
\label{ApvWRHD}
A(\Xi^- \to \Sigma ^- U^0)&= & \frac{5}{3\sqrt{2}}~s_R.
\end{eqnarray}
The factor $1/(m_i-m_f)$ in Eq.(\ref{DintermsofA}) is cancelled by factors
$m_i-m_f$ contained in amplitudes $A(B_i \to B_f U^0)$ 
since $b_R$ and $s_R$ vanish in the limit $m_i-m_f \to 0$,
as discussed after Eq.(\ref{bRcR}).

In Table \ref{bRcoefficients} we show the sizes of the coefficients at the
$b_R$ terms in Eqs (\ref{ApvWRHD}) (and hence, up to a factor, 
in Eq. (\ref{DintermsofA})) for the SU(3)-symmetric case ($x=0$,
$\epsilon =1$), and
the SU(3)-breaking case ($x=1/3$, $\epsilon=2/3$).
Please note that with growing $x$  the $\Sigma ^+ \to p \gamma$ coefficient 
 increases from zero very fast,
 so that at $x=1/3$ it becomes larger  than the absolute value
 of the corresponding coefficient
 for $\Xi^0 \to \Lambda
\gamma $
(for $\Xi ^0 \to \Lambda (\Sigma ^0) \gamma $
the relevant change is of the order of 30\%, as naively expected for
SU(3) breaking effects). Thus, for the $\Sigma ^+ \to p \gamma $ p.v. amplitude
the SU(3) breaking effect is very large indeed.

\begin{table}
\caption{Numerical values of coefficients at $b_R$ in Eqs. (\ref{ApvWRHD})}
\label{bRcoefficients}
\begin{center}
\begin{tabular}{c|c|c}
Process & $x=0$, $\epsilon =1 $ & $x=1/3$, $\epsilon =2/3$\\
\hline
$\Sigma ^+ \to p \gamma $&$0$&$0.196$\\
$\Lambda \to n \gamma $ &$0.192$&$0.048$\\
$\Xi ^0 \to \Lambda \gamma $&$-0.192$&$-0.128$\\
$\Xi ^0 \to \Sigma ^0 \gamma$ &$-0.333$&$-0.5$\\
\hline
\end{tabular}
\end{center}
\end{table}

\section{From radiative to nonleptonic decays}

Since the parity-conserving WRHD amplitudes are known via the symmetry
connection
to
the parity-conserving NLHD amplitudes (see Table \ref{tablepcWRHD}),
the branching ratios and asymmetries of WRHD provide information on the size of
parity-violating WRHD amplitudes, and, consequently,
on parameters $b_R$ and $s_R$. 

Present data on WRHD are gathered in Table \ref{pvWRHDdatafit}.
In order to get information on the size of $b_R$ and $s_R$ we performed
fits to the five known branching ratios (given in Table \ref{pvWRHDdatafit}) 
and the three well-known asymmetries (as in Table \ref{pvWRHDdatafit} 
with the exception of $\Xi^- \to \Sigma^-
\gamma$). 
Since only a rough description of the data can be achieved in this way,
we decided not to use the experimental errors in the fitting procedures.
Still, the fits yield a fairly well-defined value of $s_R$ (around $-0.75$).
For the fixed value $s_R = -0.75$ 
one can study then how the branching ratios and
asymmetries depend on the value of $b_R$.
In Table \ref{pvWRHDdatafit} we present results of such calculations for three
values of $b_R$, i.e. for $b_R=+4.2$, $+5.3$, and $+6.5$. One can see
that in this range of $b_R$ theory is in a reasonable agreement with the data.

Quantification of this agreement in terms of a $\chi ^2$-like function depends
on the details of how the errors are treated.
A reasonable requirement to impose 
is to admit equal deviations from unity of the ratios
of $x_i\equiv {\cal{B}}_i(the)/{\cal{B}}_i(exp)$ and 
$y_k \equiv \alpha _k(the)/\alpha _k(exp)$
with ${\cal{B}}_i$ ($\alpha _k$) denoting the branching ratios (asymmetries)
in question. Since the position of the minimum depends somewhat on whether one uses 
$\sum_i(x_i-1)^2+\sum_k (y_k-1)^2$ or a similar function
with $x_i \to 1/x_i$ and $y_k \to 1/y_k$,
we decided to consider the minimization of the function
\begin{equation}
\sum_{i=1}^5\left( 
\frac{{\cal{B}}_i(the)-{\cal{B}}_i(exp)}{{\cal{B}}_i(the)+{\cal{B}}_i(exp)}
\right)^2 + \sum_{k=1}^3
\left(
\frac{\alpha_k(the)-\alpha_k(exp)}{\alpha_k(the)+\alpha_k(exp)}
\right)^2,
\end{equation}
which embodies such requirements in a more symmetric way (i.e. it treats
the theoretical and the experimental entries in the same way). 
The fitted values of
$b_R$ and $s_R$ are then (in units of $10^{-7}$)
\begin{eqnarray}
b_R&\approx & +5.3\\
s_R& \approx & -0.75.
\end{eqnarray}
Putting aside the $\Xi^- \to \Sigma ^- \gamma$ branching ratio which
depends on $s_R$ only,
we observe from Table \ref{pvWRHDdatafit}
that the $\Xi^0 \to \Sigma^0 \gamma$ branching ratio is overestimated
while the branching ratios of $\Sigma ^+ \to p \gamma$, 
and $\Lambda \to n \gamma$ are underestimated.
This suggests that there might be a problem with $\Xi^0 \to \Sigma^0 \gamma$.
Consequently, it seems more likely that $b_R $ is somewhat larger than $5.3$.

For the $\Sigma ^+ \to p \gamma $ the discrepancy between the model and
experiment is about 20\% at the amplitude level.
For the $\Xi ^0 \to \Sigma ^0 \gamma $ the discrepancy is larger.
There seems to be an even larger discrepancy for 
the $\Lambda \to n \gamma $ branching ratio,
but - as already discussed - this is the decay 
for which strong cancellations occur in the
parity-conserving amplitude with the result depending quite substantially
on the precise value of 
the transition moment $\mu _{\Sigma \Lambda}$.
The overall description of the data is rough but
fairly satisfactory and indicates that
$b_R \approx -b_S$ as discussed in \cite{Zen03}.
The fits have a clear tendency to choose a small negative value for $s_R$,
thus predicting a substantial positive asymmetry for the
$\Xi ^- \to \Sigma ^- \gamma$ decay.
For comparison, in Table \ref{pvWRHDdatafit} we also
quoted the branching ratios and asymmetries 
calculated in \cite{Gav}.

\begin{table}
\caption{Fit to branching ratios and asymmetries of weak radiative hyperon
decays; data from 
\cite{PDG} and from \cite{NA482004} (marked with $^*$).}
\label{pvWRHDdatafit}
\begin{center}
\begin{tabular}{c|lcccc}

Process &Data & & Fit &  & ref.\cite{Gav}
\\
&     & $b_R=+4.2$  & $b_R\approx +5.3$  &  $b_R=+6.5$ &
\\
&     & $s_R= -0.75$ & $s_R\approx -0.75$& $s_R= -0.75$&
\\
\hline
& \multicolumn{5}{c}{Branching ratio (in units of $10^{-3}$)} 
 \\
\hline
$\Sigma ^+ \to p \gamma$ & $1.23 \pm 0.05$&$0.68$&$0.72$&$0.78$&$0.92^{+0.32}_{-0.14}$
\\
$\Lambda \to n \gamma $&$1.75 \pm 0.15 $&$0.74$&$0.77$&$0.80$& $0.62$
\\
$\Xi ^0 \to \Lambda \gamma$&$1.16 \pm 0.08^{*)}$ &$0.91$&$1.02$&$1.17$&$3.0$
\\
$\Xi ^0 \to \Sigma ^0 \gamma$&$3.33\pm 0.10$&$3.80$&$4.42$&$5.33$& $7.2$
\\
$\Xi^- \to \Sigma ^- \gamma$&$0.127 \pm0.023$&$0.16$&$0.16$&$0.16$ &$$
\\
\hline
 & \multicolumn{5}{c}{Asymmetry} 
\\
\hline
$\Sigma ^+ \to p \gamma$ & $-0.76 \pm 0.08$ &$-0.54$&$-0.67$&
$-0.79$&$-0.80^{+0.32}_{-0.19}$
\\
$\Lambda \to n \gamma $&$$&$-0.90$&$-0.93$&$-0.95$&$-0.49 $
\\
$\Xi ^0 \to \Lambda \gamma$&$-0.78\pm 0.19^{*)}$ &$-0.92$&$-0.97$&$-0.99$&$-0.78$
\\
$\Xi ^0 \to \Sigma ^0 \gamma$&$-0.63\pm 0.09$&$-0.78$&$-0.92$&$-0.99$&$-0.96$
\\
$\Xi^- \to \Sigma ^- \gamma$&$+1.0\pm 1.3$&$+0.8$&$+0.8$&$+0.8$&
\\

\hline
\end{tabular}
\end{center}
\end{table}

When one inserts the value $b_R \approx 6.0 \times 10^{-7}$ 
into Eq.(\ref{bRcR}) one obtains (in units of $10^{-7}$)
\begin{equation}
-5 \approx -12.9 +\frac{6.0}{1+x} = -8.4,
\end{equation}
or, if the estimate of $b_P$ (Eqs.(\ref{DGHfPdP},\ref{bPcPdata2})) 
from \cite{DGH} is used,
\begin{equation}
-5 \approx -11.1 +\frac{6.0}{1+x} = -6.6,
\end{equation}
where the factor $1/(1+x)$ takes into account the $SU(3)$ breaking in the
propagators of amplitudes (b2). (This is consistent with the analogous 
factors used in the
derivation of Eqs (\ref{ApvWRHD}).)
The discrepancy between the $P$- and $S$- waves is now significantly smaller,
especially for the values of $f_P$ and $d_P$ extracted in \cite{DGH}.

If one accepts that the small size of $s_R$ suggests the smallness of $c_R$
($s_R$ contains contributions from $c_R$ and diagrams (a), (a')
 in Fig.1, and therefore
one cannot determine $c_R$ uniquely), one concludes that one should have
\begin{equation}
\label{above}
c_S \approx c_P.
\end{equation}
This is indeed true for the parameters of \cite{DGH} for which
Eq.(\ref{above}) reads:
\begin{equation}
12 \times 10^{-7} \approx 13.4 \times 10^{-7}.
\end{equation}
In conclusion,we have shown
that  the argument of ref.\cite{Zen03}
works fairly well also when SU(3) 
is broken in the p.v. sector as well.
Still,
some room for unaccounted corrections is obviously present.

\section{Conclusions}

The aim of this paper was to provide both an
 explanation of the 
S:P puzzle in NLHD and a successful description of the gross structure of the
observed pattern of asymmetries and branching
ratios in WRHD in the SU(3) breaking case, the two explanations being related
as discussed
in \cite{Zen03}. The scheme
maintains an intimate connection between NLHD and WRHD, uses VMD,
and yet it does not lead to the Hara's-theorem-violating results
of both the constituent quark model \cite{KR} and 
the original VMD approach of \cite{Zen89,Zen91}. 

The resolution of the problem of weak hyperon decays given in the present 
paper was originally 
suggested in \cite{LeY,Gav}. However,
complex quark model calculations of \cite{LeY,Gav} did not make it easy
to see the simple $SU(2)_W$ symmetry connection existing 
between the p.v. WRHD amplitudes and the correction term in 
the p.v. NLHD amplitudes.
In ref. \cite{LeY}
the correction to the CA commutator term in the p.v. NLHD amplitudes
is due to the $(70,1/2^-)$ intermediate
states. In our approach
explicit calculations 
of the contributions from the individual intermediate states and 
the subsequent summation are not performed.
Instead, we work at the level of the total resulting contribution.
Still, the symmetry properties of the correction term in
ref.\cite{LeY} and in this paper are identical in the appropriate 
limit.
The difference is that in our paper, 
instead of calculating the overall size of the correction in a quark model
as the authors of ref.\cite{LeY} do, we extract both its size and sign
 from WRHD (via $SU(2)_W$ and VMD), 
 thus bypassing many quark model uncertainties.

Our identification of how symmetry should be applied for a successful
joint description of nonleptonic and radiative weak hyperon decays
leads to problems elsewhere, however.
Namely, present understanding of nuclear parity violation (cf. 
ref.\cite{DDH})
is based on symmetry between the {\em full} p.v. weak amplitudes 
$B' \to BP$ and $B' \to BV$.
According to \cite{DDH,Despl} the explanation of data on nuclear parity
violation can be obtained through the dominance of the weak
$\rho$-nucleon coupling of the form $\bar{u}_N\gamma _{\mu} \gamma _5 u_N
{\rho }^{\mu}$. Via vector-meson dominance this leads 
to photon-nucleon coupling $\bar{u}_N\gamma _{\mu} \gamma _5 u_N
{A}^{\mu}$ which entails
the violation of Hara's theorem \cite{LZ,Zen89}.
Since 
Hara's theorem is satisfied, it follows
that either the standard form of
VMD is not universal
or our present understanding of
nuclear parity violation 
is not fully correct.

{\bf Acknowledgements}\\
This work was supported in part by the
Polish State Committee for Scientific Research 
grant 2 P03B 046 25.

\vfill
\newpage

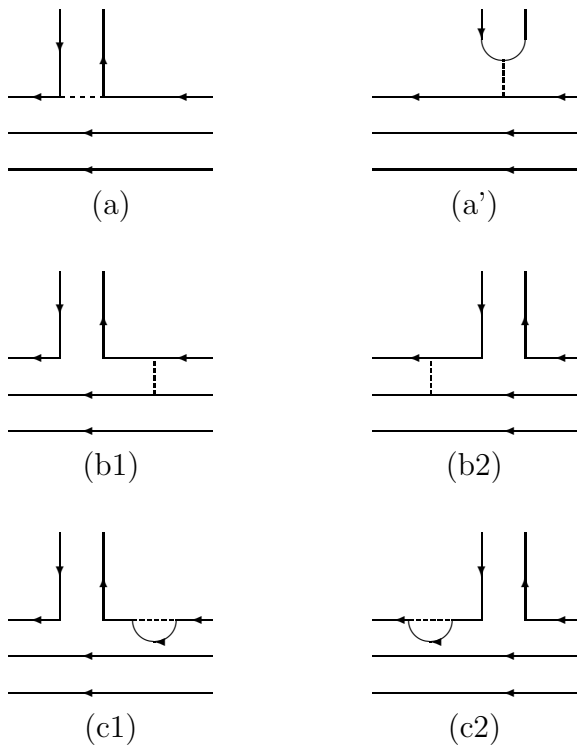
Fig.1 Quark diagrams for parity-violating weak transitions $B_i \to B_f M$

\setlength{\unitlength}{0.55pt}
%-------DIAGRAMS ---of --FIG.1
\begin{picture}(550,850)
\put(60,99){
\begin{picture}(450,710)
%---------------------------
%diagram (a1)
\put(0,550){
\begin{picture}(200,160)
\put(100,15){\makebox(0,0){(a)}}
\put(170,90){\vector(-1,0){25}}
\put(145,90){\line(-1,0){50}}
\multiput(65,90)(7,0){5}{\line(1,0){3}}
\put(65,90){\vector(-1,0){20}}
\put(45,90){\line(-1,0){15}}
\put(65,150){\vector(0,-1){30}}
\put(65,120){\line(0,-1){30}}
\put(95,90){\vector(0,1){30}}
\put(95,120){\line(0,1){30}}
\put(170,65){\vector(-1,0){90}}
\put(80,65){\line(-1,0){50}}
\put(170,40){\vector(-1,0){90}}
\put(80,40){\line(-1,0){50}}
\end{picture}}
%--------------------------
%diagram (b1)
\put(0,370){
\begin{picture}(200,160)
\put(100,15){\makebox(0,0){(b1)}}
\put(170,90){\vector(-1,0){25}}
\put(145,90){\line(-1,0){50}}
\multiput(130,65)(0,5){5}{\line(0,1){3}}
\put(65,90){\vector(-1,0){20}}
\put(45,90){\line(-1,0){15}}
\put(65,150){\vector(0,-1){30}}
\put(65,120){\line(0,-1){30}}
\put(95,90){\vector(0,1){30}}
\put(95,120){\line(0,1){30}}
\put(170,65){\vector(-1,0){90}}
\put(80,65){\line(-1,0){50}}
\put(170,40){\vector(-1,0){90}}
\put(80,40){\line(-1,0){50}}
\end{picture}}
%-----------------------------
%diagram (c1)
\put(0,190){
\begin{picture}(200,160)
\put(100,15){\makebox(0,0){(c1)}}
\put(170,90){\vector(-1,0){15}}
\put(155,90){\line(-1,0){10}}
\put(115,90){\line(-1,0){20}}
\multiput(115,90)(5,0){6}{\line(1,0){3}}
\put(130,90){\oval(30,30)[b]}
\put(130,75){\vector(-1,0){0}}
\put(65,90){\vector(-1,0){20}}
\put(45,90){\line(-1,0){15}}
\put(65,150){\vector(0,-1){30}}
\put(65,120){\line(0,-1){30}}
\put(95,90){\vector(0,1){30}}
\put(95,120){\line(0,1){30}}
\put(170,65){\vector(-1,0){90}}
\put(80,65){\line(-1,0){50}}
\put(170,40){\vector(-1,0){90}}
\put(80,40){\line(-1,0){50}}
\end{picture}}
%--------------------------
%diagram (d1)
%\put(0,10){
%\begin{picture}(200,160)
%\put(100,15){\makebox(0,0){(d1)}}
%\put(170,90){\vector(-1,0){25}}
%\put(145,90){\line(-1,0){50}}
%\multiput(130,40)(0,5){5}{\line(0,1){3}}
%\put(65,90){\vector(-1,0){20}}
%\put(45,90){\line(-1,0){15}}
%\put(65,150){\vector(0,-1){30}}
%\put(65,120){\line(0,-1){30}}
%\put(95,90){\vector(0,1){30}}
%\put(95,120){\line(0,1){30}}
%\put(170,65){\vector(-1,0){90}}
%\put(80,65){\line(-1,0){50}}
%\put(170,40){\vector(-1,0){90}}
%\put(80,40){\line(-1,0){50}}
%\end{picture}}
%--------------------------------
%diagram (a2)
\put(250,550){
\begin{picture}(200,160)
\put(100,15){\makebox(0,0){(a')}}
\put(170,90){\vector(-1,0){15}}
\put(155,90){\line(-1,0){20}}
\put(135,130){\line(0,1){20}}
\put(105,150){\vector(0,-1){20}}
\put(170,65){\vector(-1,0){50}}
\put(120,65){\line(-1,0){90}}
\put(170,40){\vector(-1,0){50}}
\put(120,40){\line(-1,0){90}}
\put(105,90){\vector(-1,0){50}}
\put(55,90){\line(-1,0){25}}
\put(105,90){\line(1,0){30}}
\multiput(120,90)(0,5){5}{\line(0,1){3}}
\put(120,130){\oval(30,30)[b]}
\end{picture}}

%--------------------------------
%diagram (b2)
\put(250,370){
\begin{picture}(200,160)
\put(100,15){\makebox(0,0){(b2)}}
\put(170,90){\vector(-1,0){15}}
\put(155,90){\line(-1,0){20}}
\put(135,90){\vector(0,1){30}}
\put(135,120){\line(0,1){30}}
\put(105,150){\vector(0,-1){30}}
\put(105,120){\line(0,-1){30}}
\put(170,65){\vector(-1,0){50}}
\put(120,65){\line(-1,0){90}}
\put(170,40){\vector(-1,0){50}}
\put(120,40){\line(-1,0){90}}
\put(105,90){\vector(-1,0){50}}
\put(55,90){\line(-1,0){25}}
\multiput(70,65)(0,5){5}{\line(0,1){3}}
\end{picture}}
%--------------------------------
%diagram (c2)
\put(250,190){
\begin{picture}(200,160)
\put(100,15){\makebox(0,0){(c2)}}
\put(170,90){\vector(-1,0){15}}
\put(155,90){\line(-1,0){20}}
\put(135,90){\vector(0,1){30}}
\put(135,120){\line(0,1){30}}
\put(105,150){\vector(0,-1){30}}
\put(105,120){\line(0,-1){30}}
\put(170,65){\vector(-1,0){50}}
\put(120,65){\line(-1,0){90}}
\put(170,40){\vector(-1,0){50}}
\put(120,40){\line(-1,0){90}}
\put(105,90){\line(-1,0){20}}
\multiput(55,90)(5,0){6}{\line(1,0){3}}

\put(55,90){\vector(-1,0){10}}
\put(45,90){\line(-1,0){15}}
\put(70,90){\oval(30,30)[b]}
\put(70,75){\vector(-1,0){0}}
\end{picture}}
%--------------------------------
%diagram (d2)
%\put(250,10){
%\begin{picture}(200,160)
%\put(100,15){\makebox(0,0){(d2)}}
%\put(170,90){\vector(-1,0){15}}
%\put(155,90){\line(-1,0){20}}
%\put(135,90){\vector(0,1){30}}
%\put(135,120){\line(0,1){30}}
%\put(105,150){\vector(0,-1){30}}
%\put(105,120){\line(0,-1){30}}
%\put(170,65){\vector(-1,0){50}}
%\put(120,65){\line(-1,0){90}}
%\put(170,40){\vector(-1,0){50}}
%\put(120,40){\line(-1,0){90}}
%\put(105,90){\vector(-1,0){50}}
%\put(55,90){\line(-1,0){25}}
%\multiput(70,40)(0,5){5}{\line(0,1){3}}
%\end{picture}}
%-----------------------------------------

\end{picture}}%-----of Fig.2
\put(300,10){\makebox(0,0)[b]{}}
\end{picture}
\end{document}